\documentclass[prd,nofootinbib,showpacs,preprint,floatfix]{revtex4}
\usepackage{latexsym}
\usepackage{dcolumn}
\RequirePackage{graphicx}

\usepackage{subfigure}

\newcommand{\MET}{{\slash\!\!\!\!E_T}}

\begin{document}

% Fix up broken revtex columns, use D{.}{.}{2} and later replace with d
%\appdef\class@documenthook{%
% \@ifpackageloaded{dcolumn}{%
%  \newcolumntype{d}{D{.}{.}{1}}%
% }{}%
%}%

\preprint{IIT-CAPP-10-01, ANL-HEP-PR-10-10}

\title{Isolated leptons from heavy flavor decays: Theory and data}

\author{Zack~Sullivan}
\email{Zack.Sullivan@IIT.edu}
\affiliation{Illinois Institute of Technology,
Chicago, Illinois 60616-3793, USA}
\author{Edmond~L.~Berger}
\email{berger@anl.gov}
\affiliation{High Energy Physics Division,
Argonne National Laboratory,
Argonne, Illinois 60439, USA}

\date{May 14, 2010}
%\date{\today}

\begin{abstract}
  Events with isolated leptons play a prominent role in signatures of
  new physics phenomena at high energy collider physics facilities.
  In earlier publications, we examine the standard model contribution
  to isolated lepton production from bottom and charm mesons and
  baryons through their semileptonic decays $b, c \rightarrow l +
  \mathrm{X}$, showing that this source can overwhelm the effects of
  other standard model processes in some kinematic domains.  In this
  paper, we show that we obtain good agreement with recent Tevatron
  collider data, both validating our simulations and showing that we
  underestimate the magnitude of the heavy-flavor contribution to the
  isolated lepton yields.  We also show that the isolation requirement
  acts as a narrow bandpass filter on the momentum of the isolated
  lepton, and we illustrate the effect of this filter on the
  background to Higgs boson observation in the dilepton mode.  We
  introduce and justify a new rule of thumb: isolated electrons and
  muons from heavy-flavor decay are produced with roughly the same
  distributions as $b$ and $c$ quarks, but with 1/200 times the rates
  of $b$ and $c$ production, respectively.
\end{abstract}


\maketitle

\section{Introduction}
\label{sec:introduction}

{\em Isolated} leptons along with missing transverse energy $\MET$ are
typical signatures for new physics processes at collider energies.  A
much-anticipated example of charged dilepton production is Higgs boson
decay, $H \rightarrow W^+ W^-$ followed by purely leptonic decays of
the $W$ intermediate vector bosons.  Charged trilepton production may
arise from the associated production of a chargino $\tilde
{\chi}_1^{\pm}$ and a neutralino $\tilde{\chi}^0_2$ in supersymmetric
(SUSY) models, followed by the leptonic decays of the chargino and
neutralino.

There are many standard model (SM) sources of isolated leptons.  The
nature and magnitude of contributions from semileptonic decays of
heavy flavors (bottom and charm quarks) are emphasized in two
papers~\cite{Sullivan:2006hb,Sullivan:2008ki}.  The role of heavy
flavor backgrounds in $H \rightarrow W^+ W^-\rightarrow l^+ l^- +
\MET$ at Fermilab Tevatron and CERN Large Hadron Collider (LHC)
energies is presented in Ref.~\cite{Sullivan:2006hb}.  We simulate the
contributions from processes with $b$ and $c$ quarks in the final
state, including $b \bar{b} X$, $c \bar{c} X$, $W c$, $W b$, $W b
\bar{b}$.  In Ref.~\cite{Sullivan:2008ki}, we study the signal and
backgrounds for $\tilde {\chi}_1^{\pm} \tilde{\chi}^0_2 \rightarrow
l^+ l^- l^{\pm} + \MET$.  We include heavy-flavor contributions to the
backgrounds from $b Z/\gamma^*$, $b \bar {b} Z/\gamma^*$, $c
Z/\gamma^*$, $c \bar {c} Z/\gamma^*$, $b \bar {b} W$, and $c \bar {c}
W$.  We learn that isolation cuts do not generally remove leptons from
heavy-flavor sources as backgrounds to multilepton searches.  A
sequence of complex physics cuts is needed, conditioned by the new
physics one is searching for.  Moreover, the heavy-flavor backgrounds
cannot be easily extrapolated from more general samples.  The
interplay between isolation and various physics cuts tends to
emphasize corners of phase space rather than the bulk characteristics.
While the heavy-flavor backgrounds can be overwhelming, we propose
specific new cuts that can help in dealing with them, and we suggest
methods for {\em in situ} verification of the background estimates.

Our finding that the dominant backgrounds to low-momentum dilepton and
trilepton signatures come from real $b$ and $c$ decays may be met with
some skepticism.  Since our publications, important Collider Detector
at Fermilab (CDF) data~\cite{Aaltonen:2008my} have appeared that allow
us to make a quantitative comparison at Tevatron collider energies.
We report the results of our comparison in this paper.  Specifically,
these data allow us to verify how well we model isolation of leptons
for events classified as originating from the ``Drell-Yan'' processes,
in which virtual photons $\gamma^*$ and intermediate vector bosons $W$
and $Z$ are produced and decay into leptons.  In addition, since we
predict the absolute rate of heavy-flavor production and the Drell-Yan
processes, we can check whether the data agree with our prediction of
the rates of isolated leptons from these sources.

In Sec.~\ref{sec:compare}, we present our detailed comparison with the
CDF data, using the same control regions defined in their study, and
using the same detector simulations and event generation methods of
our previous papers~\cite{Sullivan:2006hb,Sullivan:2008ki}.  We obtain
good agreement with the CDF data, both validating our simulations and
showing that our estimates of the magnitude of the heavy-flavor
contribution are conservative.

The added confidence in our understanding of the backgrounds from
heavy-flavor sources motivates another look at one aspect of our study
of $H \rightarrow W^+ W^-\rightarrow l^+ l^- + \MET$.  Our results
show a sharp fall of the contribution from heavy-flavor decays at
large values of the dilepton transverse mass distribution.  This
falloff is too steep to reflect only the drop with transverse momentum
of the cross section for heavy-flavor production.  In
Sec.~\ref{sec:isofilter}, we explain how isolation serves as a narrow
bandpass filter on the momentum of the leptons, thus explaining the
steep decrease at high mass.
    
A discussion of the implications of our results is found in
Sec.~\ref{sec:conclusions}.  We utilize the effect of the bandpass
filter on $b$ and $c$ decays as a way to develop a simple
rule-of-thumb that 1/200 of every produced bottom or charm quark is
seen as an isolated muon, and another 1/200 is seen as an isolated
electron, each with roughly the momentum of the heavy quark.  We end
this introduction with a general discussion of isolation and a summary
of its effects.

Given a lepton track and a cone of size $\Delta R$ in rapidity and
azimuthal angle space, the lepton is said to be {\em isolated} if the
sum of the transverse energy of all other particles within the cone is
less than a predetermined value (either a constant or a value that
scales with the transverse momentum of the lepton).  Our simulations
based on the known semileptonic decays of bottom and charm hadrons
show that leptons which satisfy isolation take a substantial fraction
of the momentum of the parent heavy hadron.  Moreover, isolation
leaves $\sim 7.5 \times 10^{-3}$ muons per parent $b$ quark.  The
potential magnitude of the background from heavy-flavor decays may be
appreciated from the fact that the inclusive $b \bar{b}$ cross section
at LHC energies is about $5 \times 10^8$~pb.  A suppression of $\sim
10^{-5}$ from isolation of two leptons still leaves a formidable rate
of isolated dileptons.  For the isolated leptons, our simulations show
that roughly $1/2$ of the events satisfy isolation because the remnant
is just outside whatever cone is used for the tracking and energy
cuts, and another $1/2$ pass because the lepton took nearly all the
energy, meaning there is nothing left to reject upon.  The latter
events are not candidates to reject with impact parameter cuts since
they tend to point to the primary vertex.  Although the decay leptons
are ``relatively'' soft, we find that their associated backgrounds
extend well into the region of new physics with relatively large mass
scales, such as a Higgs boson with mass $\sim 160$~GeV.

\section{Comparison to CDF}
\label{sec:compare}

\begin{figure*}[htb]
\centering
\includegraphics[width=\columnwidth]{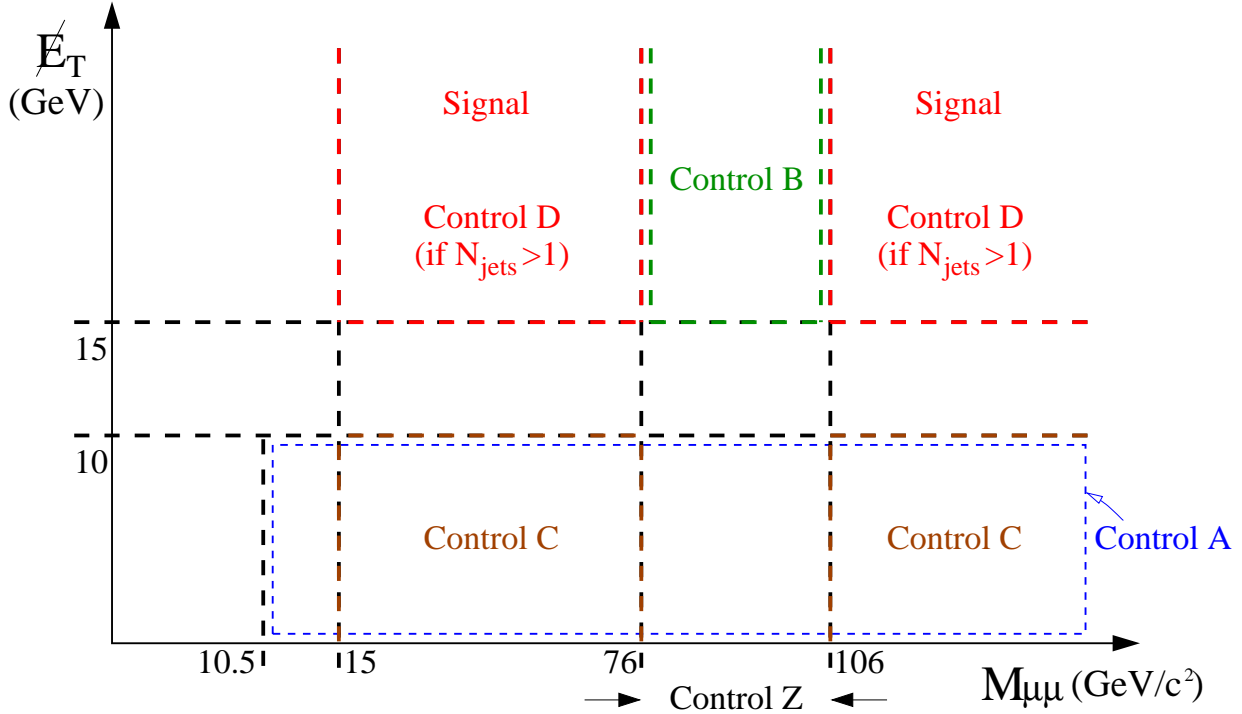}
\caption{Control regions used in the CDF analysis defined in terms of the
missing transverse energy $\MET$ vs.\ dimuon invariant mass $M_{\mu^+\mu^-}$
plane.  Figure is reproduced from Fig.\ 5 of Ref.\ 
\protect\cite{Aaltonen:2008my}.
\label{fig:regions}}
\end{figure*}

In their analysis of isolated leptons from $b$ decay
\cite{Aaltonen:2008my,CDFweb}, CDF defines several control regions in
order to disentangle the effects of different backgrounds.  In Fig.\
\ref{fig:regions}, we reproduce Fig.\ 5 of their paper
\cite{Aaltonen:2008my} displaying the various regions.  In order,
control region Z is the $Z$-boson resonance, and corresponds to an
opposite-sign dimuon invariant mass acceptance of $76 < M_{\mu^+\mu^-}
< 106$~GeV.  Region A is a low missing transverse energy region, with
$\MET < 10$~GeV and $M_{\mu^+\mu^-} > 10.5$ GeV.  The CDF region S is
designed as a signal region for their trilepton study (but examines
dimuons here) with $\MET > 15$ GeV, $\le 1$ jet, $M_{\mu^+\mu^-} >
15$~GeV, and excludes region Z.  Regions B, C, and D complement region
S: region B is a subset of region Z, with $\MET > 15$~GeV; region C is
a subset of region A with $M_{\mu^+\mu^-} > 15$~GeV and excludes the
overlap with region Z; and region D is the same as region S, but
requires at least 2 jets.

Our goal is to compare directly with the CDF measurement of isolated
leptons from $b$ decay \cite{Aaltonen:2008my,CDFweb} in each of the
control regions defined above using exactly the same detector
simulations and methods as our previous papers
\cite{Sullivan:2006hb,Sullivan:2008ki}.  We concentrate on the
contributions from $b\bar b$ pair production, with the semileptonic
decay $b\to\mu+X$, and the Drell-Yan process $p\bar p\to
\gamma^*/Z+X$.  We also consider the contributions from $W$ bosons
plus heavy flavors ($Wc$, $Wb$, $Wc+\mathrm{jet}$, $Wb+\mathrm{jet}$,
$Wb\bar b$, and $Wc\bar c$), though they contribute significantly only
to region S.  We do not include muons from light-quark jets, such as
$K$ or $\pi$ decays, which are usually classified as ``fakes.''  We
also do not predict the rate of muons from $c\bar c$, as this was not
separately identified by CDF.

\begin{figure}[htb]
\centering
\includegraphics[width=3in]{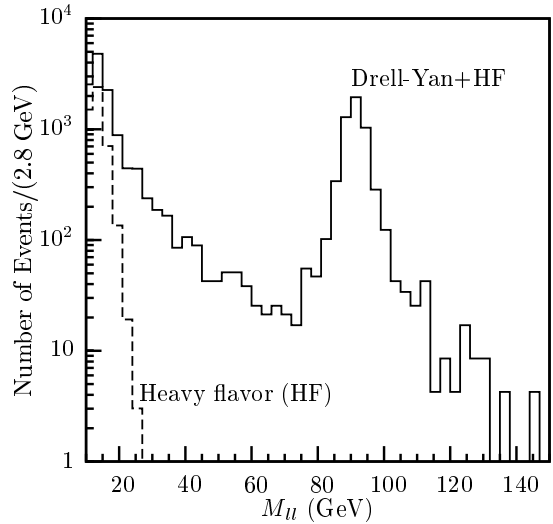}  % actual width=3.25in
\caption{Dimuon invariant mass distribution from Drell-Yan and
heavy-flavor production in region A.
\label{fig:mll}}
\end{figure}

We generate events with a customized version of MadEvent 3.0
\cite{Maltoni:2002qb} and run them through the PYTHIA 6.327
\cite{PYTHIA} showering Monte Carlo.  Both programs use the CTEQ6L1
parton distribution functions \cite{Pumplin:2002vw} evaluated via an
efficient evolution code \cite{Sullivan:2004aq}.  The showered events
are fed through a version of the PGS 3.2 \cite{Carena:2000yx} fast
detector simulation, modified to match CDF geometries, efficiencies,
and detailed reconstruction procedures \cite{Acosta:2005mu}.  At the
level of individual reconstructed leptons and jets we reproduce CDF
full detector simulations and data acceptance to a few percent.  In
Fig.\ \ref{fig:mll} we show the dilepton invariant mass distribution
in region A.  This figure compares well with Fig.\ 6c of the CDF
analysis \cite{Aaltonen:2008my}, indicating that our predictions of
the shapes agree with the data.

An important point regarding our predictions is that they are
absolutely normalized.  We apply $K$-factors to the leading order
rates of $1.4$ for Drell-Yan, and $1.4$ for $b\bar b$
production\footnote{Note, this factor is smaller than the $K$ factor
  of $2.0$ typically assumed for $b\bar b$, and that would be obtained
  for a different choice of scale, and a more inclusive measurement.}.
These $K$-factors are calculated via the NLO program MCFM \cite{MCFM}
including cuts that mimic the final sample in order to more accurately
predict the result in the region of CDF acceptance.

In order to compare to the CDF data, we scale our results to those
observed in region Z, the $Z$-peak, where we do not expect any signal
from $b\bar b$ events.  Theoretical errors due to parton distribution
functions and Monte Carlo statistics are included, but they are
smaller than the error propagated from the experimental measurement of
region Z used to normalize the data.  We see in
Table~\ref{tab:results} that with this scaling we reproduce well the
results for Drell-Yan production in control regions A and C --- the
regions with the most statistical significance.

\begin{table}[tbh]
\caption{Comparison between the number of the Drell-Yan (DY) and $b\bar b$
events observed by CDF (from Table III of Ref.\ 
\protect\cite{Aaltonen:2008my}) in each control region, and our predictions.
Dashes indicate no events are expected or observed.
($^*$) We could not obtain enough statistics in Control D to compare.
($^\dag$) Includes $10\pm 3$ events from $W+\mathrm{heavy\ flavors}$
(mostly $Wc$).
\label{tab:results}}
\begin{tabular}{ccc|cc}
&\multicolumn{2}{c}{CDF} & \multicolumn{2}{|c}{Our study}\\[-0.5ex]
Region & DY & $b\bar b$ & DY & $b\bar b$\\\hline
Control Z & $6419\pm 709$ & --- & $6419\pm 752$ & ---\\
Control A & $14820\pm 2242$ & $9344\pm 1621$ &
 $14222\pm 1615$& $5118\pm 584$\\
Control B & $217 \pm 25$ & --- & $58.9 \pm 24.9$ & ---\\
Control C & $5770\pm 1043$ & $2238\pm 384$ &
 $4898\pm 584$ & $924\pm 117$\\
Control D & $7.8 \pm 1.5$ & $9 \pm 4$ & $9.8 \pm 9.9$ & ---$^*$\\
Control S & $169 \pm 30$ & $90 \pm 20$ & $226 \pm 53.2$ & $26 \pm 10^\dag$
\end{tabular}
\end{table}

Once we have normalized to the observed luminosity, we find that our
predictions systematically \textit{underestimate} the number of
isolated leptons from $b$ decays in the overlapping control regions (A
and C) by nearly a factor of two.  This result suggests that our
analysis has been conservative in estimating this generally ignored
source of contamination.  Part of the observed difference may be due
to the small $K$-factor of $1.4$ we use to estimate the absolute
normalization.  However, even a $K$-factor of 2 would leave a
systematic underestimate of $1.4$.  The remainder could be due to
contamination from $c\bar c$ decays that produce isolated muons on one
side of the event, and resemble $b$'s on the other side.  Regardless,
the background is not only a large fraction of the entire data set,
but it is more significant than the estimations of Refs.\
\cite{Sullivan:2006hb,Sullivan:2008ki}.

In our main analysis, we do not consider regions D and S of the CDF
study significant for two reasons.  First, the statistics are too
small to make a definitive statement.  Second, the choice of regions
is sensitive to the reconstruction of missing transverse energy
$\MET$, which is difficult to model at the level of 10--15 GeV.  One
concern may be that events have ``slipped'' from the low $\MET$
regions to the high ones.  If this were due to a systematic shift in
our reconstruction it should also appear in the Drell-Yan sample.  We
see a hint of this in our Drell-Yan estimate for region S.  However,
we see the opposite effect in isolated leptons from heavy-flavor
decays, where our prediction of 16 events from $b\bar b$ and 10 events
from $W+\mathrm{heavy\ flavors}$, underestimates the measured CDF
region by a factor of 3.5.\footnote{Using the lower estimate of 70
  events from the CDF measurement and our upper estimate of 36 events
  leads to the same factor of 2 under-prediction observed in other
  regions.}  The underestimate of isolated leptons from heavy-flavor
decays in region S is consistent with our reduced estimate of
Drell-Yan in region B.

Given that region Z overlaps both regions C and S, a comparison that
should reduce the sensitivity to $\MET$ would be between the sums of
these regions.  In that case, our Drell-Yan result (C$+$S) of $5124
\pm 586$ events does agree a bit better with the CDF observation of
$5939 \pm 1043$, but in both cases is within $1\sigma$.  If we
consider isolated leptons from heavy-flavor decays, we find regions
(C$+$S) have $950 \pm 120$ events, which is still approximately a
factor of two smaller than the combined CDF regions of $2328 \pm 385$.
Hence, it appears unlikely that misestimations of $\MET$ or
$W+\mathrm{heavy\ flavor}$ decays, are responsible for our systematic
underestimate of the heavy-flavor background.

Figure \ref{fig:mll} demonstrates that the bulk of the CDF dimuon
sample from heavy flavors is composed of muons with transverse
momentum $p_T$ less than 10 GeV.  One limitation of our study is that
our original detector simulation was constructed and tuned for leptons
with $p_T > 10$--20 GeV.  It is somewhat surprising we are able to
model the detector response as well as we do.  Perhaps the $Z$ region,
where we trust our detector simulation, is not representative enough
of the low dilepton invariant-mass region.  To this end, we also
consider in Table~\ref{tab:region} normalizing our $b\bar b$
contribution in each region independently.  We do this by extracting
the $K$-factor necessary to exactly scale our Drell-Yan calculation to
the measured Drell-Yan rate, and applying that to our prediction of
the $b\bar b$ contamination. Under this method, our predicted
background from $b\bar b$ increases by less than one standard
deviation, and it is still about a factor of 2 smaller than CDF data.

\begin{table}[tbh]
\caption{Comparison between the number of $b\bar b$ events observed in
CDF control regions, and our predictions, where each region is normalized
separately by the ratio of the CDF measurement of Drell-Yan production
in that region divided by our prediction.
($^\dag$) Includes $8\pm 3$ events from $W+\mathrm{heavy\ flavors}$
(mostly $Wc$).
\label{tab:region}}
\begin{tabular}{ccc}
Region & CDF & Our study\\\hline
Control A & $9344\pm 1621$ & $5333 \pm 833$\\
Control C & $2238\pm 384$ & $1089 \pm 213$\\
Control S & $90 \pm 20$ & $20 \pm 8^\dag$
\end{tabular}
\end{table}

We conclude this section with the statement that our estimate of the
contamination to isolated lepton sample from heavy-flavor decays
appears conservative.

\section{Lepton isolation acts as a momentum filter}
\label{sec:isofilter}

\begin{figure}[htb]
\centering
\includegraphics[width=3in]{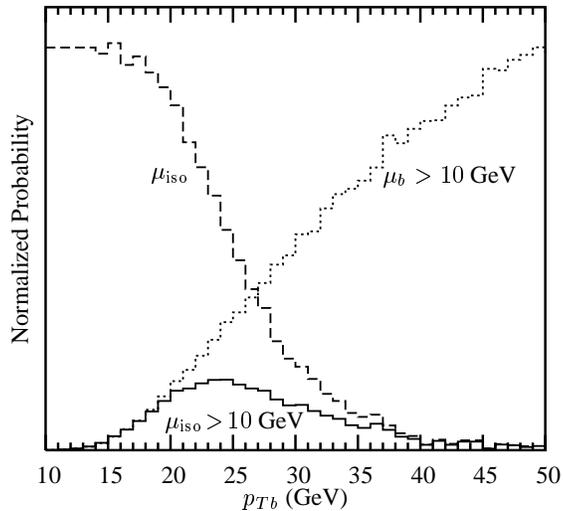}  % actual width=3.25in
\caption{Normalized probability for a $b$ quark to produce an isolated muon
with $p_{T\mu}>10$ GeV (solid) vs.\ the $b$ transverse momentum
\cite{Sullivan:2008ki}.  This curve is a multiplicative combination of the
probability of producing a muon with $p_{T\mu}>10$ GeV (dotted) and the
probability the muon will be isolated (dashed).  The $b$ production
spectrum is not included.  
\label{fig:muvptb}}
\end{figure}

Having experimental verification that isolated leptons from heavy
flavor decays play an important role, we clarify one aspect that
experimental reconstruction has on the spectrum of these leptons.  In
Sec.\ II of Ref.\ \cite{Sullivan:2008ki}, we explain in detail how
leptons from $b$ and $c$ decays pass isolation cuts.  Briefly, the
probability to produce a lepton above some threshold, e.g., $p_{T\mu}
> 10$ GeV is convolved with the probability of missing the rest of the
$b$ or $c$ decay remnant.  For completeness, we reproduce the shape of
this efficiency using an ATLAS-like detector simulation for muons as a
function of the $p_{Tb}$ of the quark in Fig.\ \ref{fig:muvptb}.

\begin{figure}[htb]
\centering
\includegraphics[width=3in]{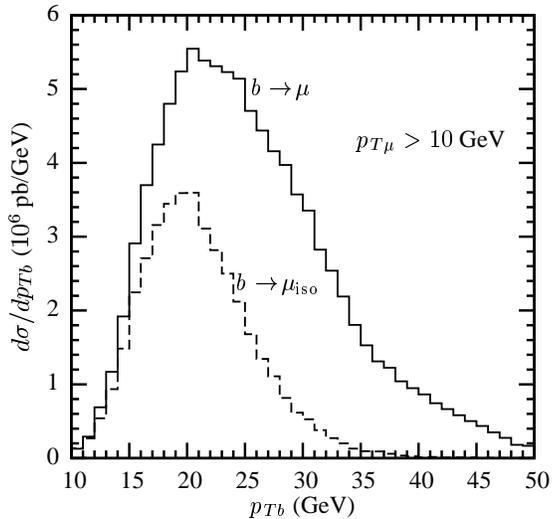}  % actual width=3.25in
\caption{Cross section for production of a muon of $P_{T\mu} > 10$~GeV
from $b\bar b$ production and decay (solid), or an isolated muon (dashed).
\label{fig:muvbb}}
\end{figure}

When the acceptance function is folded with a typical $b$ transverse
momentum spectrum --- whether from $b\bar b$ production, or any other
spectrum that falls with $p_{T}$ --- the peak of the resulting cross
section of isolated muons comes from $b$'s whose $p_T$ is just above
the muon $p_T$.  This effect is clearly present in Fig.\
\ref{fig:muvbb}, where the peak for muons with $p_{T\mu} > 10$ GeV in
$b\bar b$ production comes from 20 GeV $b$'s.  Previously
\cite{Sullivan:2008ki}, we focused on the low-momentum end of this
spectrum, noting that a significant fraction of isolated leptons arise
from $b$'s just above threshold for production.  Hence, to properly
model this background, $b$ and $c$ must be modeled all the way down to
threshold.

In this section, we focus on the upper end of the isolated lepton
spectrum as a function of $b$ (or $c$) transverse momentum.  The long
tail of acceptance in Fig.\ \ref{fig:muvptb} is suppressed by the
sharply falling $b$ transverse momentum spectrum.  The net result is
that the isolation acts as a \textit{narrow bandpass filter on
momentum}.  This observation has a critical importance for
higher-scale physics, such as Higgs boson production.

\begin{figure*}[tbh]
\centering
\subfigure[]{\includegraphics[width=3in]{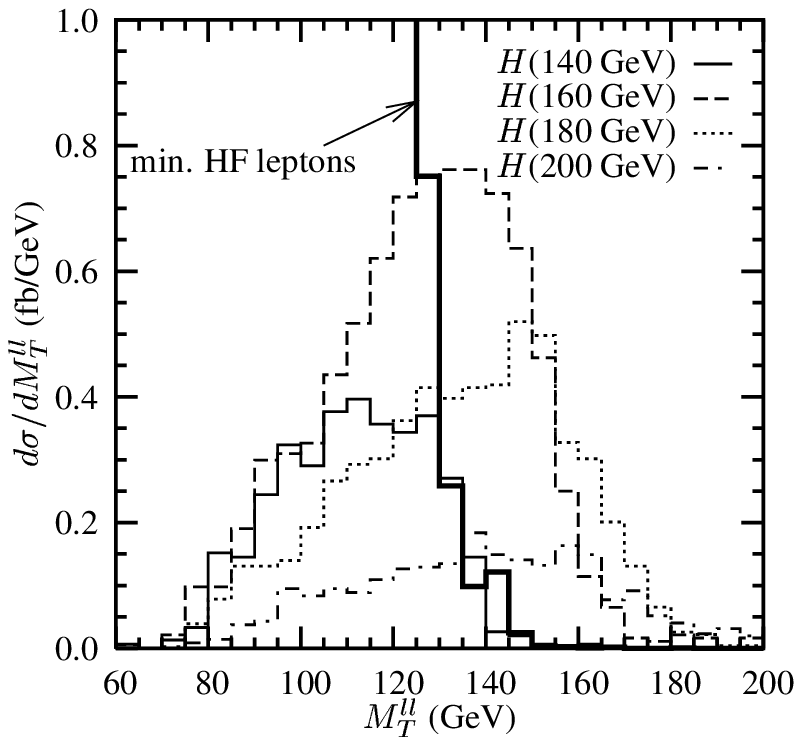}}
\subfigure[]{\includegraphics[width=3in]{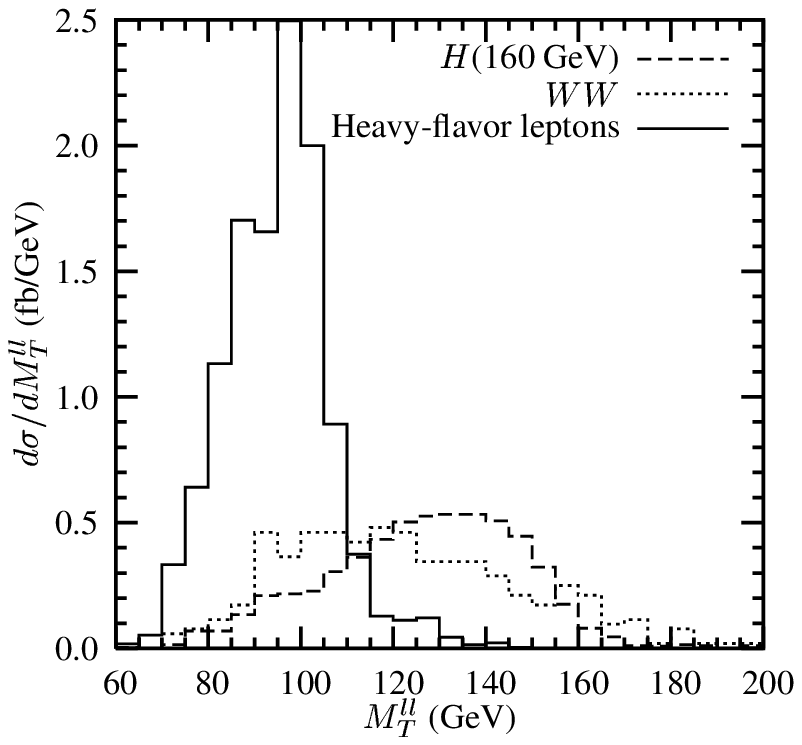}}  % actual width=3.25in
\caption{(a) Opposite-sign dilepton transverse mass for various Higgs
boson masses at $\sqrt{S}=14$~TeV, and the additional heavy-flavor
background (thick solid line), with default ATLAS cuts.
(b) Opposite-sign dilepton transverse mass for a 160 GeV Higgs boson, the
continuum $WW$ background, and the additional heavy-flavor background
after the $p_{Tl_2}$ threshold of the \textit{second}-highest transverse
momentum lepton is raised from 10 GeV to 20 GeV \cite{Sullivan:2006hb}.
\label{fig:mthmm}}
\end{figure*}

In the dilepton analysis of a Higgs boson decaying to $W^+W^-$ to
dileptons we observe that the transverse-mass $M_T$ spectrum due to
$b\bar b$/$c\bar c$ and $W+$heavy flavors drops sharply at large
$M_T$.  In Fig.\ \ref{fig:mthmm}(a) the background from heavy flavors
falls through the middle of Higgs boson signals for masses in the
range 140--200 GeV.  Raising the $p_T$ threshold of the
\textit{second-highest $p_T$ lepton} from 10 GeV to 20 GeV pushes this
leading edge to lower transverse mass, Fig.\ \ref{fig:mthmm}(b),
thereby recovering our ability to extract a Higgs signal in the
dilepton channel despite the heavy-flavor background.

Considering lepton isolation criteria as a \textit{narrow bandpass
  filter}, we see that the sharp high-$M_T$ edge is due to the cutoff
of large-momentum $b$'s by this filtering mechanism.  Hence, even
though we are modeling the tail of a steeply falling spectrum, the
filtering effect of lepton isolation acts to safely suppress the
region where we expect our $M_T$ shapes to be less-well defined.  The
net result of this filter is to provide a more robust determination of
the shape of the background from isolated leptons.

\section{Discussion}
\label{sec:conclusions}

In this paper we predict the rate of isolated muons from b decay in
order to compare directly with data from the CDF Collaboration.  The
central conclusion to draw is that isolated leptons from $b$ decays
are a large fraction of the low transverse momentum lepton sample.  In
the case of the CDF measurement we \textit{under-predict} the measured
rate of dimuons from $b$ decay using the same codes and procedures we
use to estimate the background to dilepton and trilepton signatures at
the Tevatron and LHC \cite{Sullivan:2006hb,Sullivan:2008ki}.  Hence,
we are confident that these backgrounds will play an important role in
the extraction of Higgs boson decays to $WW$, trilepton supersymmetry,
and indeed all processes with any low transverse momentum electron or
muon.

Given the broad nature of our conclusions, and the significant
computing resources required to model these backgrounds properly, we
introduce a new ``rule-of-thumb'' to determine whether these leptons
may be problematic in any given analysis:
\begin{itemize}
\item Replace 1/200 of every produced $b$ or $c$ quark with a muon,
  and 1/200 with an electron having the same momentum as the $b$ or $c$.
\item If the resulting background is more than 10\% of the signal, it
  should be simulated more carefully, and eventually measured \textit{in situ}.
\end{itemize}
This new rule-of-thumb works precisely because the lepton isolation
criteria act as a bandpass filter selecting leptons from $b$ or $c$
quarks whose transverse momenta are only slightly above the momentum
of the lepton.  In general, this rule-of-thumb is valid for large
transverse momentum leptons as well.  Fortunately, the production
rates for $b$ and $c$ quarks tend to fall rapidly (with the exception
of $b$ decays from top quarks, which peak near 50--60 GeV).  Overall,
we strongly recommend that all analyses involving leptons consider the
background from the decay of heavy-flavor hadrons, as many analyses
are sensitive to regions of phase space in which this background is
enhanced.

\begin{acknowledgments}
  E.\ L.\ B.\ is supported by the U.~S.\ Department of Energy under
  Contract No.\ DE-AC02-06CH11357.  We gratefully acknowledge the use
  of JAZZ, a 350-node computer cluster operated by the Mathematics and
  Computer Science Division at Argonne as part of the Laboratory
  Computing Resource Center.  We wish to thank John Strologas for
  discussions regarding details of the CDF data.
\end{acknowledgments}

\end{document}